\documentclass{article}

\usepackage{graphicx}
\usepackage{epsfig}
\usepackage{amssymb}
\usepackage{amsmath}
\usepackage{amsfonts, latexsym}
\usepackage{stmaryrd}
\usepackage[numbers,sort&compress]{natbib}

\newcommand{\be}{\begin{equation}}

\newcommand{\ee}{\end{equation}}
\newcommand{\ba}{\begin{array}}
\newcommand{\ea}{\end{array}}
\newcommand{\bea}{\begin{eqnarray}}
\newcommand{\eea}{\end{eqnarray}}
\newcommand{\bi}{\begin{itemize}}
\newcommand{\ei}{\end{itemize}}

\newcommand{\cH}{{\mathcal {H }}}

\newtheorem{theorem}{Theorem}

\newtheorem{prop}{Proposition}
\newtheorem{corollary}{Corollary}

\usepackage{authblk}

\begin{document}

\title{A family of fourth-order superintegable systems with rational potentials related to Painlev\'e VI}
\author{I. Marquette}
\affil[1]{School of Mathematics and Physics\\
 The University of Queensland,\\
  St Lucia, Brisbane, Queensland, 4072, AUSTRALIA}
\author{ S. Post}
\author{L. Ritter}
\affil[2]{
Department of Mathematics\\
University of Hawai`i at M\-{a}noa\\
2565 McCarthy Mall\\
Honolulu, HI 96815, USA }

\maketitle
\abstract{ We discuss a family of Hamiltonians given by particular rational extensions of the singular oscillator in two-dimensions. The wave functions of these Hamiltonians can be expressed in terms of  products of Laguerre and exceptional Jacobi polynomials. We show that these systems are superintegrable and admit an integral of motion that is of fourth-order. As such systems have been classified, we see that these potential satisfy a non-linear equation related to Painlev\'e VI. We begin by demonstrating the process with the simpler example of rational extensions of the harmonic oscillator and use the classification of third-order superintegrable systems to connect these families with the known solutions of Painlev\'e IV associated with Hermite polynomials.  }

\section{Introduction}
The connection between classical orthogonal polynomials and superintegrable systems has long been the subject of various research programs.  A guiding principle is that superintegrable systems are those that can be solved algebraically as well as analytically and it is these properties that characterize the classical orthogonal polynomials. More precisely, if the system admits separation of variables then its eigenfunctions will be solutions of a Sturm-Liouville equation and the additional integrals will give recurrence relations between the energy eigenstates. 

Looking beyond the well-studied systems with second-order superintegrability (or multiseparability) in two-dimensions, several authors have investigated superintegrable systems that admit separation of variables, implying a second-order integral of motion, as well as an additional higher-order integral of motion \cite{GW, Gravel, MW2008, TW20101, PopperPostWint2012, Marchesiello2015,  MSW2017, Escobar-Ruiz2018, Escobar-Ruiz2018General}. For our purposes, we'd like to highlight the work of Gravel \cite{Gravel} who investigated systems that separate in Cartesian coordinates and admit third-order integrals of motion. Later, Tremblay and Winternitz \cite{TW20101} investigated systems that separate in polar coordinates with integrals of third-order and   Escobar-Ruiz,  L\'opez Vieyra  and Winternitz\cite{Escobar-Ruiz2017, Escobar-Ruiz2018, Escobar-Ruiz2018General} extended the investigation to fourth-order and beyond. A striking result of these investigations is that while the systems themselves are linear, the determining equations for the existence of these higher-order integrals are non-linear resulting in potentials that depend on solutions of nonlinear equations. In particular, the potentials for  the quantum Hamiltonian systems have been associated with Painlev\'e equations or higher-order differential equations with the Painlev\'e property. 

On the flip side of the research done on classifying superintegrable systems is the wide range of results on the construction of superintegrable systems with integrals of higher-order \cite{TTW, PW2010, marquette2011infinite, ranada2012new,  ranada2013higher, Marquette2009}. One method of this construction is to begin with systems whose wavefunctions are given by special functions or orthogonal polynomials and to use the properties of these functions to construct the integrals of the motion \cite{Marquette2009a, Marquette2010,Marquette2010a, KKMTools2010, KKMRecurrence2011}. For example, several such systems have been constructed using exceptional orthogonal polynomials as wavefunctions \cite{quesne2008exceptional, quesne2011higher, PostVinet2012, marquette2013new}. Exceptional orthogonal polynomials are orthogonal polynomials that also satisfy a Sturm-Liouville equation except with gaps in their degree sequence \cite{Gomez-Ullate2009, Gomez-UllateDavid2010}. This puts them outside of the classification of Bochner and allows for rational coefficients in the eigenvalue equations. 

These rational extension of known exactly-solvable or superintegrable systems have been investigated by several authors and have been linked to potential associated with Painlev\'e transcendents \cite{Odake2009, Odake2011, Marquette2013, Marquette2016, Fernandez2016, Gomez-Ullate2020}. In a previous paper by the authors \cite{MPR2020}, a system associated with exceptional Jacobi polynomials was constructed, shown to be superintegrable and identified in the classification by Escobar-Ruiz, L\'opez Vieyra and Winternitz \cite{Escobar-Ruiz2017} to be associated with Painlev\'e VI. In this paper, we consider an entire family of such systems, show that they are all superintegrable and all belong to the same classification. Thus, we have an infinite family of rational extensions of the trigonometric Rosen–Morse potential potential that satisfy a non-linear equation equivalent to Painlev\'e VI. We also discuss the algebra and representations of these systems. Before presenting the main results, we review a similar construction for the rational extension of the harmonic oscillator, where the details are much simpler, to elucidate the main ideas behind the analysis. 

For the remainder of this introduction, we briefly review the results in Ref \cite{MPR2020}. We begin with the following Hamiltonian
\begin{equation} H = -\frac{1}{2} \Delta + \frac{1}{2} \omega^2 r^{2} + \frac{1}{2 r^{2}}  \left( \frac{ \alpha^{2} - \frac{1}{4}}{\sin^{2}{\theta}} + \frac{\beta^{2} -\frac{1}{4} }{\cos^{2}{\theta}} + \frac{8(1+ \gamma \cos{(2 \theta}))}{(\gamma + \cos{2 \theta})^{2}} \right),
\end{equation}
with $\gamma = \frac{\alpha - \beta}{\alpha + \beta}$. This Hamiltonian was introduced in \cite{PostVinet2012} in a family of superintegrable systems as an exceptional deformation of the TTW system \cite{TTW}. In particular, the members of the family depended on an index $k$ which was multiplied by the variable $\theta$ and  which determined the order of the additional integrals of the motion. In Ref. \cite{MPR2020}, we focused on the $k=1$ case and verified that the system was superintegrable. We showed that there were no third-order integrals of motion and gave an explicit representation of a fourth-order integral 
\begin{eqnarray}
	L=  (p_1^{2} -p_2^{2} )M^2+ \{ g(x,y), p_1^{2} \}  +\{ h(x,y), p_1p_2\} +\{ i(x,y)  p_2^2\} + k(x,y),  \label{L}
\end{eqnarray}
with the functions $g, h, i$, and $k$ defined in the paper. Here $p_i$ and the momenta in Cartesian coordinates and $M =\partial_\theta$ is the generator of rotations. 

Comparing our results to classification work on superintegrable sytems that separate in polar coordinates, we verified that the system has no third-order integrals and so is outside of the third-order classification \cite{MW2008}. Turning to the classification of such systems with fourth-order integrals of motion\cite{Escobar-Ruiz2017},  we see that any potential that admits a fourth-order integral of the form (\ref{L}) and can be expressed as 
 \[ V=\frac{1}{2}\omega^2 r^2+\frac{2}{r^2}T'(\theta)\] 
will satisfy a non-linear equation equivalent to equation SD-I.a in the classification by Cosgrove \cite{Cosgrove2000a}. In particular, we take
\[ T(\theta)=\frac{-2 W(\theta)}{\sin(\theta)\cos(\theta)} -\frac{\alpha^2-\alpha\beta +\beta^2+\frac{7}{4}}{\tan(2\theta)},\]
and making the change of variables $y= \frac12 (1+\cos(2\theta))$ giving the following nonlinear equation for $W(y):$
\[ y^2(1-y)^2(W")^2+4W'(yW'-W)^2-4(W')^2(yW'-W)
\]
\begin{equation}\label{SD1}
+4q_7(W')^2 +4q_8W'+4q_9(yW'-W)+4q_{10}=0,\end{equation}
for certain constants $q_i$. In particular, the solutions to this equation are expressible in terms of Painlev\'e transcendents. Thus, we have a link between this rational extension of this singular ocillator/Rosen-Morse potential and Painlev\'e VI. The purpose of this current work is to extend these results to rational extensions of arbitrary order, thus creating an infinite family of rational functions related to PVI transcendents.

\section{Rational Extensions of the 2D Harmonic Oscillator and PIV}
Let us begin with a straightforward example of the process. We first consider rational extensions of the two-dimensional harmonic oscillator connected with exceptional Hermite polynomials\cite{Marquette2013, Gomez-Ullate2014, Marquette2016}. The one-dimensional  harmonic oscillator can be expressed in terms of ladder operators 
\be a= \dfrac{d}{dx}+x, \qquad a^{\dagger} =-\dfrac{d}{dx} +x,\ee
which are mutual adjoints with respect to the Euclidean inner-product 
\[ \langle f , g \rangle= \int_{-\infty}^{\infty} f\,  \overline{g}\,dx.\]
The ladder operators satisfy the Weyl algebra relations
\[[a,a^\dagger] =2.\]
The standard harmonic oscillator is created from these ladder operators as 
\be \label{Hx} H_x= aa ^{\dagger}-1,\ee
\[H_x = -\dfrac{d^2}{dx^2}+x^2.\]
The eigenfunctions of this operator are given by 
\[ \Psi_n(x)= e^{-x^2/2} H_n(x), \qquad H_x \Psi_n(x)= (2n+1)\Psi_n(x), \]
where $H_n(x)$ are the Hermite polynomials defined as 
\[ H_n(x)= (2x)^n {}_2F_{0}\left( \begin{array}{cc} -n/2, & -(n-1)/2\\ -&\end{array}\bigg| \frac{-1}{x^2}\right). \]

The two-dimensional system is then constructed in Cartesian coordinates as the direct sum of these two Hamiltonians
\[ H =H_{x}+H_y,\]
with a complete eigenbasis of wavefunctions given by
\[ \Psi_{n,m}(x,y) = e^{-(x^2+y^2)/2} H_n(x)H_m(y), \qquad H\Psi_{n,m}(x,y) = (2n+2m+2)\Psi_{n,m}(x,y).\]
It is clear that each energy level has a set of degenerate eigenvalues (hidden symmetry) with $n+m=N$. Two linearly independent, self-adjoint integrals of motion can be created by exploiting this degeneracy as
\[ L_1 =\frac{i}{2}\left( a^{\dagger}_xa_y- a^{\dagger}_y a_x \right),\]
\[ L_2 =\frac{1}{2}\left( a^{\dagger}_xa_y+ a^{\dagger}_y a_x \right).\]
In coordinates, the integrals take the following form
\[ L_1 =i\left(x\dfrac{\partial}{\partial y} -y \dfrac{\partial }{\partial x}\right), \qquad L_2= - \dfrac{\partial^2}{\partial x\partial y} +xy.\]

This method can be extended to rational extensions of the harmonic oscillator by making use of their ladder operators. Let us consider just the state adding approach and the corresponding factorization of the Hamiltonian as in \cite{Marquette2016}. A one-step rational extension can be constructed based on the following factorization 
\[ A =\dfrac{d}{dx} +W(x), \qquad A^\dagger = -\dfrac{d}{dx}+W(x),\]
\[ W(x) = -x - \frac{\cH_k'(x)}{\cH_k(x)},\]
where $\cH_k(x)$ is a psuedo-Hermite polynomial 
\[ \cH_k(x) =(i)^{-k}H_k(ix)\] 
and $k$ is assumed to be even so that there are no real zeros for these seed functions. 

The rational extension of the 1D harmonic oscillator is then given by 
\[ H^{(2)}=AA^\dagger -1,\]
and satisfies 
\[ AH=H^{(2)}A, \qquad A^\dagger H^{(2)}=HA^\dagger.\]
Here the superscript denotes the fact that this would be a one-step extension and the procedure could be applied iteratively. We will not consider such extensions here.

The (unnormalized) eigenfunctions of this rationally extended Hamiltonian are given in terms of the exceptional orthogonal polynomials, 
\[ H^{(2)} \psi_n^{(k)}(x) =(2n+1) \psi_n^{(k)}(x), \qquad n=1, k+1, k+2, \ldots  \]
\[ \psi_n^{(k)}(x) = \frac{e^{-x^2/2}}{\cH_k(x)} y_n^{(k)}(x), \]
\[ y_n^{(k)}= \left\{ \begin{array}{lr} 1 \qquad &n=0\\
-\cH_k H_{n-k} -2k\cH_{k-1} H_{n-k+1}, \qquad &n\geq k
\end{array}\right.\]

Using these expressions for the wave-functions,  combined with the ladder operators for the Hermite polynomials, we obtain ladder operators for this exceptional basis via
\[ b^\dagger = A a^\dagger A^\dagger, \qquad b= A a A^\dagger,\]
which satisfy 
\[ [ H^{(2)}, b^\dagger]= 2b^\dagger, \qquad [ H^{(2)}, b]=2b.\]
Thus, we can directly verify that a rationally extended 2D Hamiltonian
\be H =H_{x}^{(2)}+H_y,\label{Hrational} \ee
will have integrals of motion of a similar form 
\begin{equation} \label{L1f} L_1 =\frac{i}{2}\left( b^{\dagger}_xa_y- a^{\dagger}_y b_x \right),\end{equation}

\begin{equation} L_2 =\frac{1}{2}\left( b^{\dagger}_xa_y+ a^{\dagger}_y b_x \right).\end{equation}

Note that,  because the ladder operators for the rational extension are of third-order, the integrals of motion $L_1, L_2$ will be of third- and fourth- order respectively. This general procedure can be extended to multiple-step extensions but here we are particularly in search of potentials that admit third-order integrals and fourth-order integrals and so confine our discussion to this regime. 

\begin{prop}
For each $k$, the Hamiltonian in (\ref{Hrational}) is a superintegrable extension of the harmonic oscillator whose potential is rational of degree $2k.$ 
 The Hamiltonian in (\ref{Hrational}) is separable in Cartesian coordinates and admits a second-order integral of motion in a single variable. It also admits a third-order integral of motion of the form 
\be \label{L1} L_1 = \{ (x \partial_y-y\partial_x), \partial_x^2\} + \{ g_1(x,y), \partial_x\} +\{ g_2(x,y), \partial_y\} .\ee
\end{prop}
The proceeding discussion describes the construction of the integrals of motion and it remains only to show by direct computation that the leading order-terms of $L_1$ (\ref{L1}).

\begin{theorem}[Gravel \cite{Gravel}]
If a Hamiltonian is separable in Cartesian coordinates and admits an integral of the form in Proposition 1 (\ref{L1}), then the potential is of the form 
\be V = y^{2} + x^{2} - g' + g ^{2} + 2 x g - 1, \label{Vg} \ee
where $g$ is a solution to Painlev\'e IV (PIV). 
\end{theorem}
We can see the proof of this theorem from the classification results of Gravel \cite{Gravel}. To give a survey of the method, a third-order integral is assume of the appropriate form. The commutation relations lead to a set of determining equations which can be solved modulo a non-linear compatibility equation for the potential. This equation was identified by Cosgrove \cite{Cosgrove2000} to be expressible in the given form in terms of PIV transcendents. Please note, we have scaled out some of the parameters from the original paper for simplicity of the expressions. 

This result indicates that there is a connection between these rational extension of the harmonic oscillator and solution of PIV. Comparing the potential from (\ref{Vg}) to the factorized form (\ref{Hx}) we can identify the functions $g(x)= {\cH_k'}/{\cH_k}$ as possible  rational solutions to PIV. A direct calculation, using identities of Hermite polynomials verifies this well known fact that rational solutions of PIV can be constructed from Hermite polynomials \cite{Marquette2016}. 

To review, we have constructed a family of superintegrable systems of rational extension to the harmonic oscillator. We have used raising and lowering operators to construct integrals of motion with the same leading term for all members of the family. We then show that the potentials associated with  superintegrable Hamiltonians with such integrals of the motion satisfy nonlinear equations whose solutions can be expressed in terms of PIV transcendents. From the form of the potential, we are able to identify the parts of the potential that give rational solutions of PIV.

\section{Rational Extension of the trigonometric Rosen–Morse potential and PVI}
We next take this method and repeat it on rational extensions of a trigonometric Rosen–Morse potential, one of the systems in the Smorodinsky-Winternitz classification. The system we begin with is separable in polar coordinates of the form 
\begin{equation}
		H_{RM} = -\frac{1}{2} \partial_{r} r \partial_{r}+ \frac{1}{2} \omega^{2} r^{2}  + \frac{1}{2r}\left( -\partial_{\theta}^{2} + \frac{\alpha^2-\frac1 4}{\cos^2(\theta)} +\frac{\beta^2-\frac1 4}{\sin^2(\theta)}
\right). \label{nonexcepH}
	\end{equation}
	The angular part of the Hamiltonian (itself an integral of the motion)
	\[ L=  -\partial_{\theta}^{2} + \frac{\alpha^2-\frac14}{\cos^2(\theta)} +\frac{\beta^2-\frac14 }{\sin^2(\theta)}\]
	can be expressed in factorized form using 
	\be a= \dfrac{d}{d\theta}+(\alpha-\beta+1)\cot(\theta) -(\alpha+1/2)\sec\theta \csc \theta,\ee
	\be a^{\dagger} =-\dfrac{d}{d\theta}+(\alpha-\beta+1)\cot(\theta) -(\alpha+1/2)\sec\theta \csc \theta,\ee
which again are mutual adjoints with respect to the Euclidean inner-product.
The angular part of the Hamiltonian can be expressed in terms of  these ladder operators as 
\be \label{L2} L= aa ^{\dagger}-4\alpha+4\beta.\ee
The eigenfunctions of this operator can be written in terms of polynomial solutions in the variable
\be x=-\cos(2\theta) \label{xdef}\ee
and are given by 
\[ \Psi_n(x)= (x+1)^{\beta/2+1/4}(x-1)^{\alpha/2+1/4}P_n^{(\alpha, \beta)}(x), \qquad L \Psi_n(x)= (2n+\alpha +\beta +1)^2\Psi_n(x), \]
where $P_n^{\alpha, \beta}(x)$ are the Jacobi polynomials defined as 
\[ P_n^{(\alpha, \beta)}(x)= \frac{(\alpha+1)_n}{n!}{}_2F_{1}\left( \begin{array}{cc} -n, & n+\alpha+\beta +1\\ \alpha+1&\end{array}\bigg| \frac{1-x}{2}\right). \]
	
Next, we give an alternative factorization for this angular piece. We continue our discussion for the remainder of the section in the variable $x$ (\ref{xdef}). The exceptional ladder operators are 
\[ A =2\sqrt{1-x^2}\left(\dfrac{d}{dx} +W(x)\right), \qquad A^\dagger = 2\sqrt{1-x^2}\left(-\dfrac{d}{dx}+W(x)\right),\]
with
\be \label{seed}W=\dfrac{d}{dx} \ln \left((x+1)^{\beta/2-1/4}(x-1)^{-\alpha/2-1/4}P_{
m}^{(-\alpha-1, \beta-1)}(x)\right) .\ee
The function $W$ is sometimes referred to as the seed-function. 
The original operator $L$ can be expressed as 
	\be \label{L2factored} L(\alpha+1, \beta-1)= A^\dagger A(\alpha, \beta)+(\alpha-\beta-1)^2
 ,\ee
	where the parameters have been shifted appropriately. 
	In the opposite factorization 
	\be \label{L2hat} L^{(2)}= AA^\dagger (\alpha, \beta)+ (\alpha-\beta-1)^2,\ee
	we obtain a rational extension of this potential. The extension gives 
	\begin{eqnarray}
	 L^{(2)}& =&4(1-x^2) \frac{d^2}{dx^2} +4x\frac{d}{dx} + \frac{4\beta^2-1}{2(1+x)}-\frac{4\alpha^2+1}{2(x-1)} \nonumber\\
	 &&+\frac{4(x^2-1) (P''_{m}(x)P_{m}-2(P'_{m}(x))^2) +4\left((\alpha-\beta+2)x +\alpha+\beta \right)P'_{m}(x)P_{m}(x)}{P_{m}(x)^2},\end{eqnarray}	
	 where the superscripts of the Jacobi polynomials have been omited as in  $P_{m}(x) =P_{
m}^{(-\alpha-1, \beta-1)}(x).$
	
	The new operator $L^{(2)}$ intertwines with the old as in the previous section 
	\be \label{intertwine} AL=L^{(2)}A, \qquad A^\dagger L^{(2)}=LA^\dagger,\ee
	and we can use the eigenfunctions for $L$ to obtain eigenfunctions for $L^{(2)}$ via
	\be \label{hatP} \hat{ \Psi}_n(x)= A\left(1-x)^{\alpha/2+3/4}(x+1)^{\beta/2-1/4}(P_n^{(\alpha+1, \beta-1)}(x)\right), \ee
	where $n=0,1, 2, \ldots$ but written in factorized form with a weight gives 
	\[ \hat{ \Psi}_n(x)=\frac{(1-x)^{\alpha/2+1/4} (x+1)^{\beta/2+1/4}}{P_{m}^{(-\alpha-1, \beta-1)}(x)}\hat{P}_{n}^{(\alpha, \beta)}(x)\]
	where $\hat{P}_{n}^{(\alpha, \beta)}(x)$ is an exceptional Jacobi polynomial of degree $n+1, $ they can be expressed using (\ref{hatP}) as 
	\[  \hat{P}_{n}^{(\alpha, \beta)}(x) =2(1-x)W\left(P_{m}^{(-\alpha-1, \beta-1)}(x), P_n^{(\alpha+1, \beta-1)}(x)\right) -2(\alpha+1)P_{m}^{(-\alpha-1, \beta-1)}(x)P_n^{(\alpha+1, \beta-1)},\]
	where $W(f,g) =f'g-g'f$ is the Wronskian of two functions. 
	
Due to the intertwining properties of the exceptional operator (\ref{intertwine}), the exceptional wave functions  $\hat{ \Psi}_n(x)$ will have the same eigenvalues as their non-exceptional counterparts 
\[ L^{(2)} \hat{ \Psi}_n(x)=(2n+\alpha +\beta +1)^2\hat{ \Psi}_n(x), \qquad n= 0,1, 2, \ldots\]
although the degree of the exceptional polynomials will be $n+m.$

We then replace the angular part of the non-exceptional Hamiltonian with this exceptional operator to obtain the rational extension
\begin{equation}
		H = -\frac{1}{2} \partial_{r} r \partial_{r}+ \frac{1}{2} \omega^{2} r^{2}  + \frac{1}{2r}L^{(2)}. \label{excepH}
	\end{equation}
	The radial part of the Hamiltonian  is as in the non-exceptional case
	\begin{equation}
	\label{radial} \left( - \frac{1}{2r} \partial_{r} r \partial_{r} + \frac{1}{2} \omega^2 r^{2} + \frac{\Lambda^{2}}{2r^{2}} \right) R(r) -ER(r)=0
\end{equation}
 is solved by the following change of variable and conjugation by the ground state
\begin{equation}
y={\omega}r^2, \qquad 
R_0(y)= y^{\Lambda/2}e^{-y/2}. \end{equation}
Resulting in the following eigenvalue equation for Laguerre polynomials
\begin{equation}
\left[y\frac{d^2}{dy^2} +(\Lambda+1-y)\frac{d}{dy}+m\right] F(y)=0,
\end{equation}
where $m$ is associated with energy quantization 
\[ E= \omega(2m+\Lambda+1),\] 
and the separation constant is found above,
\[\Lambda= 2n+\alpha+\beta+1.\]
The radial part of the eigenfuctions then become
\[ R_{m,\Lambda}(r)=  r^{\Lambda}e^{-\omega r^2/2} L_m^{(\Lambda)}(\omega r^2),\qquad  m=0,1,2, \ldots.\]

In summary, we have wave functions for the Hamiltonian
\[ H\Phi_{m,n}(r, \theta) = E_{m,n} \Phi_{m,n}(r, \theta) \]
with separated wave functions
\[ \Phi_{m,n}(r, \theta)  = \hat{ \Psi}_n(x) R_{m, \Lambda}(r)\]
and quantized energy 
\[ E_{m,n}= \omega(2m+2n+\alpha+\beta +2 ).\] 

\subsection{Building integrals of the motion}
Although the hidden symmetry of this system is more complex, due to the separation in polar coordinates, we can still follow a similar procedure as before and use ladder operators for the separated solutions to build integrals of motion for the system. 
Let us first mention that the operator $L^{(2)}$ is an integral of the motion. For the the remainder of the paper, we will refer to this operator as $L_2$ indicating that it is a second-order integral of the motion. The remaining integrals will be labeled similarly.

We begin with raising/lowering operators for the radial parts of the Hamiltonian. These are based on the ladder operators for Laguerre polynomials conjugated by the square root of the weight to match our radial functions $R_{m}^{\Lambda}(y)$. These ladder operators can be expressed most succinctly using the quantized energy as 
\be \label{ccd} C_{\Lambda, E} = \frac{(\Lambda+1)}{r} \dfrac{\partial }{\partial r} -\frac{\Lambda(\Lambda+1)}{r^2} +E, \qquad C_{\Lambda, E}^\dagger = C_{-\Lambda, E}.\ee
These have the following action on the separated functions 
\[ C_{\Lambda, E}R_{m, \Lambda}(r) =2\omega R_{m-1, \Lambda+2}(r), \qquad C^\dagger_{\Lambda, E}R_{m, \Lambda}(r) =2\omega(m+1)(m+\Lambda) R_{m+1, \Lambda-2}(r).\]

For the ladder operators for the angular part, we begin with ladder operators for our non-exceptional wave functions, i.e. classical Jacobi polynomials multiplied by the square root of the weight. These are 
\be  b_{\Lambda}=2(1-\Lambda)(1-x^2)\dfrac{\partial}{\partial x} +\Lambda(1-\Lambda)x -(\alpha+\beta)(\alpha-\beta+2) \qquad b_{\Lambda}^\dagger=b_{-\Lambda}.\ee
They have the following action on the separated functions 
\[b_{\Lambda}\Psi_{n}(x) = -4(n+\beta-1)(n+\alpha+1)\Psi_{n-1}(x)\]
\[ b^{\dagger}_{\Lambda}\Psi_n(x)=-4(n+1)(n+\alpha+\beta+1)\Psi_{n+1}(x).\]
We can then construct the ladder operators for the angular part of the wave function by including the factorization operators 
\[ B_{\Lambda}(x)= A b_{\Lambda} A^{\dagger}, \qquad B^\dagger_{\Lambda}=B_{-\Lambda},\]
where the second equality follows from the expression for $b^\dagger_{\Lambda}$ and from the fact that the factorization operators do not depend on the angular quantum number $\Lambda.$

We can then build operators that depend on the quantities $\Lambda$, $E$ but fix the energy as follows 
\[ \Xi_{-} = B_{\Lambda} C_{-\Lambda,E}, \qquad  \Xi_{+} = B_{-\Lambda} C_{\Lambda,E}\]
with action 
\[ \Xi_{+}\Psi_{n}(x) R_{m, \Lambda}(r) =-8\omega(n+\beta-1)(n+\alpha+1)(2n+\alpha +\beta +1)^2\Psi_{n+1}R_{m-1, \Lambda+2}(r)\]
\[\Xi_{-}\Psi_{n}(x) R_{m, \Lambda}(r) =-8\omega(n+1)(n+\alpha+\beta+1)(2n+\alpha +\beta +1)^2(m+1)(m+2n+\alpha+\beta+1)\Psi_{n-1}R_{m+1, \Lambda-2}(r).\]
Here, we recall that $\Lambda= 2n+\alpha +\beta +1$ so $\Lambda+2$ corresponds with $n+1$ and similarly for $n-1.$ 

To obtain integrals of the motion that are independent of the quantities $\Lambda$ and $E$ we will replace $\Lambda^2$ by $L_2$ and $E$ by $H$. We first ensure that that the relevant quantities are polynomial in these variables. Indeed, using the parity in $\Lambda$ we can see that 
\[ L_5= \frac{1}{\Lambda} \left(B_{\Lambda} C_{-\Lambda}- B_{-\Lambda} C_{\Lambda}\right)\]
and 
\[ L_6 =\left(B_{\Lambda} C_{-\Lambda}+B_{-\Lambda} C_{\Lambda}\right).\]
will be even polynomials in $\Lambda$.  The subscripts indicate the degree of the resulting operators. However, as indicated in our previous paper\cite{Marquette2020}, for at least one choice of seed function (in particular with $m=1$),  we have a fourth-order integral. Following the methodology set out in \cite{Kalnins2011}, we look for a linear combination 
\[ L_4 = c_1 B_{\Lambda} C_{-\Lambda} +c_2 B_{-\Lambda} C_{\Lambda}+c_3\]
which satisfies $[L_2,L_4]=L_5$ on the basis  (thus determining $c_1$ and $c_2$) and introduce the additive constant $c_3$ to ensure that the resulting quantity is a polynomial in $A$. 
First we notice that $L_2 \Phi_{m,n}= \Lambda^2 \Phi_{m,n},$ and hence we can compute immediately
\be \label{XiL2} [ L_2,\Xi_{+}] =4(1+\Lambda ) \Xi_{+}, \qquad [L_2, \Xi_{-}] =4(1-\Lambda) \Xi_{-}.\ee
Thus, to create an operator $L_4$ that satisfies $[L_4, L_2]=L_5$ on the basis  we take 
\[ L_4 = \frac{1}{4\Lambda(1-\Lambda)}  B_{\Lambda} C_{-\Lambda} - \frac{1}{4\Lambda(1+\Lambda)} B_{-\Lambda} C_{\Lambda}+c_3\]
This operator has the required action on the basis and we could continue the process further, dividing by more factors to obtain successive solutions of the inverse commutator relationship. However, we need to ensure that the resulting operator is a polynomial in $\Lambda^2,$ and in particular we have not created any poles. We will begin by determining the coefficients $D, \, \tilde{D}$
 \[ K= \frac{1}{4(1-\Lambda)}  B_{\Lambda} C_{-\Lambda} - \frac{1}{4(1+\Lambda)} B_{-\Lambda} C_{\Lambda}+\frac{D}{(1-\Lambda)}+\frac{\tilde{D}}{1+\Lambda}\]
such that $K$ is an odd polynomial in $\Lambda$ and then we will obtain the required operator by dividing by $\Lambda.$
The conditions for removal of the poles are that 
\[ D=\frac{-1}{4}  B_{\Lambda} C_{-\Lambda}|_{\lambda=1} = \frac{-(\alpha+\beta)(\alpha-\beta+2)}{4}AA^\dagger E,\]
\[\tilde{D}=\frac{1}{4}  B_{-\Lambda} C_{\Lambda}|_{\lambda=-1} \]
and hence $\tilde{D}=-D.$ The resulting operator 
 \[ K= \frac{1}{4(1-\Lambda)}  B_{\Lambda} C_{-\Lambda} - \frac{1}{4(1+\Lambda)} B_{-\Lambda} C_{\Lambda}- \frac{(\alpha+\beta)(\alpha-\beta+2)\Lambda}{2(1-\Lambda^2)}AA^\dagger E\]
is an odd polynomial in $\Lambda$ and so our $L_4$, and even polynomial in $\Lambda$ is given by
 \[ L_4= \frac{1}{4\Lambda(1-\Lambda)}  B_{\Lambda} C_{-\Lambda} - \frac{1}{4\Lambda(1+\Lambda)} B_{-\Lambda} C_{\Lambda}- \frac{(\alpha+\beta)(\alpha-\beta+2)}{2(1-\Lambda^2)}AA^\dagger E.\]
 We can then accomplish the replacements $E\rightarrow H$ and $\Lambda^2 \rightarrow L_2$ to obtain a differential operator that will commute with the Hamiltonian. 
 We complete the section be determining the leading order terms of this integral. Recall, from the construction that the operator $L_4$ will be equal to the coefficients of the odd terms in $K$ (divided by $\Lambda$) and this will be 2 times the odd terms of $\frac{1}{4(1-\Lambda)}  B_{\Lambda} C_{-\Lambda}$, ignore terms with poles as they are canceled by the last term. We then investigate this quantity, ignoring the constant factor of 1/2 since we are only interested in the form of the leading order terms. The operator can be expressed as 
 \be \frac{B_{\Lambda} C_{-\Lambda}}{(1-\Lambda)}=A\left( 2(1-x^2)\dfrac{\partial}{\partial x} +\Lambda x -\frac{(\alpha+\beta)(\alpha-\beta+2)}{1-\Lambda}\right)A^\dagger  C_{-\Lambda}
 \ee
 
 We take each term separately and separate only the odd parts
 \[A\left( 2(1-x^2)\dfrac{\partial}{\partial x}\right)A^\dagger C_{-\Lambda} =\Lambda A\left( 2(1-x^2)\dfrac{\partial}{\partial x}\right)A^\dagger\left( \frac{1}{r} \dfrac{\partial }{\partial r} -\frac{1}{r^2}\right) + \mathfrak{F} (1, \Lambda^2)\]
 \[A\Lambda x A^\dagger  C_{-\Lambda}=\Lambda AxA^\dagger\left(\frac{1}{r} \dfrac{\partial }{\partial r} -\frac{\Lambda^2}{r^2} +E\right)+ \mathfrak{F} (1, \Lambda^2).\]
 For the final term we obtain
 \[A\frac{-(\alpha+\beta)(\alpha-\beta+2)}{1-\Lambda}A^\dagger C_{-\Lambda}=-(\alpha+\beta)(\alpha-\beta+2)AA^\dagger \left( \frac{1}{r} \dfrac{\partial }{\partial r} -\frac{\Lambda}{r^2} +\frac{E}{1-\Lambda}\right). \]
 The pole at $\Lambda=1$ has been removed and so the only odd part of this term is 
 \[ \frac{-\Lambda(\alpha+\beta)(\alpha-\beta+2)}{r^2} AA^\dagger.\]
 We are now prepared to express the operator $L_4$, including replacing $\Lambda^2$ and $E$ with their corresponding differential operators 
 \be \label{L4final} 2L_4= A\left( 2(1-x^2)\dfrac{\partial}{\partial x}\right)A^\dagger\left( \frac{1}{r} \dfrac{\partial }{\partial r} -\frac{1}{r^2}\right)+ AxA^\dagger\left(\frac{1}{r} \dfrac{\partial }{\partial r} -\frac{L_2}{r^2} +H\right)-\frac{(\alpha+\beta)(\alpha-\beta+2)}{r^2} AA^\dagger.\ee
From this expression,  we see that we have a fourth-order integral of motion whose leading-order term can be computed in polar coordinates using 
\[ A =-\dfrac{\partial}{\partial \theta}+\ldots,\]
\[2(1-x^2)\dfrac{\partial}{\partial x}=\sin(2\theta) \dfrac{\partial}{\partial \theta},\]
and 
\[ H-\frac{L_2}{r^2} =-\frac12 \dfrac{\partial^2}{\partial r^2}+\frac{1}{2r^2} \dfrac{\partial^2}{\partial r^2}\ldots.\]
The leading order term is then 
\[ L_4 =\frac12 \left(-\cos(2\theta)\dfrac{\partial^2}{\partial r^2}+2\sin(2\theta) \dfrac{\partial^2}{\partial r\partial \theta}+\frac{\cos(2\theta)}{r^2}\dfrac{\partial^2}{\partial \theta^2}\right)\dfrac{\partial^2}{\partial \theta^2}+\ldots.\]
In particular, we see that the leading order term is independent of the seed function. 

Let us collect the preceding discussion into a proposition. 
\begin{prop}
For any integer value $m$, the Hamiltonian (\ref{excepH}) associated with the seed function 
\[W=\frac{\left(P_{m}^{-\alpha-1, \beta-1}(x)\right)'}{P_{m}^{-\alpha-1, \beta-1}(x)}\]
will be superintegrable. It admits a second-order integral $L_2$ associated with separation of variables in polar coordinates and a fourth-order integral (\ref{L4final}) with leading order term  proportional to $(p_1^2-p_2)^2M^2. $
\end{prop}
Thus, we can extend our discussion from the $m=1$ case \cite{Marquette2020} to the entire family of exceptional Hamiltonians considered. We state the results as a corollary.
\begin{corollary}
For any integer value $m$, the potential of the Hamiltonian (\ref{excepH}) associated with the seed function $W$ (as in Proposition 2) can be expressed in terms of solutions of $PVI.$
\end{corollary}
As discussed in the introduction, this is a direct application of the classification work in \cite{Escobar-Ruiz2017}.

	\section{Cubic Algebra and Structure Functions}
We next turn to the algebra generated by the integrals of motion for this system. Using Mathematica, we were able to express the following commutators as polynomials in the generators. Here we emphasize that we used a differential equation satisfied by $W(x)$, rather than the expression in (\ref{seed}), in order to obtain the algebra coefficients for all potentials in our infinite family. 

The algebra relations are 
	\begin{eqnarray*}
	\left[ L_{2},L_{4} \right] &=& L_{5} \\
	\left[ L_{2},L_{5} \right] &=& a L_{2}^{2} + b \{L_{2},L_{4}\} + c L_{2} + d L_{4} + f \\
	\left[ L_{4},L_{5}\right]  &=& g L_{2}^{3} + h L_{2}^{2} - b L_{4}^{2} - a \{L_{2},L_{4}\} + i L_{1}- c L_{4} +j,
\end{eqnarray*}
with 
\begin{eqnarray*}
&a = 0, \quad b = 8, \quad c =\frac{4(-2\alpha+\alpha^2 +2\beta-\beta^2)}{\omega}H, \\
&d=-16, f=-\frac{4(\alpha-\beta)(-2+\alpha+\beta)(-1+\alpha+\beta+2 m_1)^2}{\omega}H, \quad g=-2,  \\
&h= 2(-1+3\alpha^2 +3\beta^2 +6\beta(-1+m)+6(-1+m)m+3\alpha(-2 +\beta+2m))+\frac{3}{2 \omega^2} H^2   \\
&i=  (2 (-3 \alpha^4 - (-1 + \beta)^2 (-5 + 3 (-2 + \beta) \beta) - 4 (1 + \beta (5 + 3 (-3 + \beta) \beta)) m\\
&- 4 (1 + 5 (-2 + \beta) \beta) m^2 - 16 (-1 + \beta) m^3 - 8 m^4 - 6 \alpha^3 (-2 + \beta + 2 m) \\
&	- 2 \alpha^2 (5 + 3 \beta^2 + 10 \beta (-1 + m) + 2 m(-9 + 5 m)) - 2 \alpha (2 + 3 \beta^3 + 10 \beta^2 (-1 + m)\\
& + 10 m+  4 m^2 (-5 + 2 m) + \beta (7 + 4 m(-7 + 3 m))))) \\
&		+ (\frac{-3+8 \beta + 8 m-4(\alpha^2 +\beta^2 +2\beta m+ 2 m^2       +\alpha(-2+\beta+2 m))}{\omega^2} H^2 \\
& j= (2 (-1 + \alpha + \beta + 2 m)^2 (\alpha^4 + \alpha^3 (-4 + \beta + 2 m) +  \alpha (4 + (-3 + \beta) \beta - 2 m) (1 + \beta + 2 m)\\
&+ 2 \alpha^2 (1 + \beta (-1 + m) + (-3 + m) m) + (-1 + \beta)^2 (-3 + \beta^2 + 2 \beta (-1 + m) + 2 (-1 + m) m)))\\
&  + ( \frac{1}{2\omega^2}(-1+\alpha+\beta+2m)^2  + (3 + 5 \alpha^2 + 5 (-2 + \beta)   \beta - 4 m+ 4 \beta m+ 4 m^2 + 2 \alpha (-5 + \beta + 2 m))  ) H^2.
\end{eqnarray*}

The Casimir operator in the differential representation takes the following values:
\begin{eqnarray*}	
& K=	\frac{1}{\omega^2}((-1 + \alpha + \beta + 2 m)^2) + (-3 + 2 \alpha^4 + 2 \beta^4 + 8 \beta^3 (-1 + m) \\
&- 20 m+ 20 m^2 + 4 \alpha^3 (-2 + \beta + 2 m) + 2 \alpha (-7 + 2 \beta^3 + \beta (13 - 16 m) \\
&+ 4 \beta^2 (-2 + m) + 18 m- 8 m^2) + \beta^2 (15 - 24 m+ 8 m^2) + \alpha^2 (15 + 4 \beta^2 \\
&+ 8 \beta (-2 + m) - 24 m+ 8 m^2) -  2 \beta (7 - 18 m+ 8 m^2)) H^2  \\
&	+(-1 + \alpha + \beta + 2 m)^2 (\alpha^6 + 2 \alpha^5 (-3 + \beta + 2 m) - \alpha^4 (-15 + \beta^2 + \beta (6 - 4 m) \\
&+ 20 m- 4 m^2) - 4 \alpha^3 (5 - \beta + \beta^3 - 10 m+ 4 m^2 +  \beta^2 (-3 + 2 m)) \\
& + (-1 + \beta)^2 (-15 + \beta^4 +  4 \beta^3 (-1 + m) - 4 m+ 4 m^2 + 2 \beta^2 (3 - 6 m+ 2 m^2) \\
&- 4 \beta (1 - 3 m+ 2 m^2)) +  2 \alpha (13 + 2 \beta^3 + \beta^5 + 10 m- 8 m^2 +   \beta^4 (-3 + 2 m) \\
&+ \beta (5 - 16 m^2) +  2 \beta^2 (-1 - 6 m+ 4 m^2)) -  \alpha^2 (1 + \beta^4 + 40 m- 24 m^2 \\
& + 4 \beta^3 (-3 + 2 m) + \beta (4 + 24 m- 16 m^2) +  \beta^2 (22 - 40 m+ 8 m^2))) 
\end{eqnarray*}	
	
Using these algebra relations, we can build a deformed oscillator representation using generators $b, b^\dagger$ and $N$ which satisfy
	\[ [N,b]=-b,\quad  [N,b^{\dagger}]=b^{\dagger} \]
\[ b^{\dagger}b=\Phi(N),\quad bb^{\dagger}=\Phi(N+1) \]
where $\Phi(N)$ is the structure function. 

The generators of our algebra are then expressed as 
\[ X=X(N), \qquad Y= Y_0(N) +b^\dagger \rho(N)+\rho(N)b\]
	with
	\[ X(N)= 4 (N+u)^2 \]
	\[  Y(N) =\frac{1}{4(-1+4(N+u)^2)\omega}(  (\alpha - \beta) (-2 + \alpha + \beta) H (-1 + \alpha + \beta - 2 N - 2 u + 2 m) (-1 + \alpha + \beta + 2 N + 2 u + 2 m) )  \]
	\[  \rho(N) = \frac{1}{33554432 \sqrt{3} \sqrt{(N+u)(1+N+u)(1+2(N+u))^2}} ,\]
	
and
\begin{eqnarray*}	
& \Phi(N) = (-4 \alpha^3 + \alpha^4 - 4 \beta^3 + \beta^4 + 4 \alpha (-2 \beta + \beta^2 + (1 - (N+u))^2) \\
&+ 4 \beta (1 - 2  (N+u))^2 + \beta^2 (2 + 8  (N+u) - 8  (N+u)^2) +  \alpha^2 (2 + 4 \beta - 2 \beta^2 + 8  (N+u) - 8  (N+u)^2)\\
& + (1 - 2  (N+u))^2 (-3 - 4  (N+u) + 4  (N+u)^2)) (-h^2 + (1 - 2  (N+u))^2 \omega^2) (-3 + \alpha^4 + \beta^4 + 8  (N+u) + 8  (N+u)^2 \\
&-  32  (N+u)^3 + 16  (N+u)^4 + 8 m- 32  (N+u) m+ 32  (N+u)^2 m+   8 m^2 + 32  (N+u) m^2\\
&- 32  (N+u)^2 m^2 - 32 m^3 +  16 m^4 + 4 \alpha^3 (-1 + \beta + 2 m) + \beta^3 (-4 + 8 m)\\
& +  4 \beta (-1 + 2 m) (-1 + 4  (N+u) - 4  (N+u)^2 - 4 m+ 4 m^2) +  \beta^2 (2 + 8  (N+u) - 8  (N+u)^2\\
& - 24 m+ 24 m^2) +  2 \alpha^2 (1 + 3 \beta^2 + 4  (N+u) - 4  (N+u)^2 - 12 m+ 12 m^2 + \\
&   6 \beta (-1 + 2 m)) + 4 \alpha (-1 + \beta + 2 m) (-1 + \beta^2 + 4  (N+u) - 4  (N+u)^2 - 4 m+ 4 m^2 + \beta (-2 + 4 m)))
\end{eqnarray*}	
Re-expressed in terms of the original Hamiltonian, the structure function factors nicely as
\begin{eqnarray*}	
&\Phi(N)= (-13510798882111488) ( (N+u - \frac{1}{2} (3 - \alpha - \beta)) (N+u -\frac{1}{2} (1 + \alpha - \beta)) \\
&(N+u -  \frac{1}{2} (1 - \alpha + \beta)) (N+u - \frac{1}{2} (-1 + \alpha + \beta)) (N+u - \frac{(-H + \omega)}{(2 \omega)}) \\
&(N+u - \frac{( H + \omega)}{(2 \omega)}) (N+u -  \frac{1}{2} (1 - \alpha - \beta - 2 m)) \\
&(N+u - \frac{1}{2} (3 - \alpha - \beta - 2 m)) (N+u - \frac{1}{2} (-1 + \alpha + \beta + 2 m))\\
&(N+u -  \frac{1}{2} (1 + \alpha + \beta + 2 m))).
\end{eqnarray*}	

Then given the action of $N$ as $N |n,E\rangle = n |n,E \rangle$, which is the energy dependent Fock space, we can compute  the action of the Casimir on the basis $K |n,k\rangle = k |n,k\rangle$, written in terms of the Hamiltonian. Then the structure functions is a function of $n$, $u$ and $E$, i.e. $\Phi=\Phi(n,u,E)$. The finite dimensional unitary representations will be connected with
\[ \Phi(0,u,E)=0,\quad \Phi(p+1,u,E)=0,\quad \Phi(x,u,E) > 0 \forall x \in [1,...,p]. \]
Assuming a vacuum state, $\Phi(0,u,E)=0,$ provide set of solutions for $u$ and we choose 
\[ u= \frac{1}{2}(-1 +\alpha + \beta ). \]
The assumption that the representation be finite leads to  $\Phi(p+1,u,E)=0,$ this giving a quantized energy value  
\[ E= \omega ( 2p + \alpha + \beta ) \]
which agrees with the energy values from the previous sections via the identification, $p=m+n$  assuming $p=1,2,3,...$ and 
$n=0,1,2,....$ and $m=0,1,2,3,...$. Other solutions depending on parameters $m$ would not be physical in our context and not be associated with square integrable states.

\section{Conclusions}
In this paper, we have analyzed two families of superintegragrable systems associated with exceptional orthogonal polynomials. In each case the systems were separable in one orthogonal coordinate system and admitted an additional, higher-order integral of motion. As such, these systems fall into the classification work of Winternitz and collaborators \cite{GW, Gravel, MW2008, TW20101, PopperPostWint2012, Marchesiello2015,  MSW2017, Escobar-Ruiz2018, Escobar-Ruiz2018General}. It is from this classification work that we are able to identify these family of potentials with solutions of non-linear equations. 
	
	We began with a family of rational extensions of the harmonic oscillator whose wavefunctions can be expressed in terms of exceptional Hermite polynomials. These systems admit a third-order integral of motion and belong to the classification of Gravel \cite{Gravel}. From this classification, we can see that the rational extensions are solutions to Painlev\'e IV, as is indeed well known there is such a family associated with Hermite polynomials\cite{Gomez-Ullate2020}. 
	
	While the method applied to the system in Section 2 recovered only known results, we then apply a similar strategy albeit with more complicated details, to a new family of rational extension of the singular oscillator. The wavefunctions of this family involve exceptional Jacobi polynomials. We use the recurrence relation method of Kalnins, Kress and Miller \cite{Kalnins2011} to show that each member of this family is superintegral and the give an explicit form for the fourth-order integral of motion. In this case, we refer to the classification of  Escobar-Ruiz, López Vieyra, and Winterntiz \cite{Escobar-Ruiz2017}, to show that these rational potentials satisfy a non-linear equation that can be transformed into Painlev\'e VI. Hence, we have obtained a novel connection between rational functions comprised of Jacobi polynomails and the sixth Painlev\'e equation. 
	
	We plan on continuing the investigation by making the connection more precise. That is,  we look to directly express a rational solution of PVI using these Jacobi polynomials. In addition, we have given algebraic relations satisfied by the integrals of motion along with deformed oscillator representation thereof. We plan to further investigate these algebra, their representations and connections to Painlev\'e equations.

\noindent{\bf Acknowledgements:} IM was supported by by Australian Research Council Future Fellowship FT180100099. SP acknowledges  The Simons Foundation Collaboration grant \# 3192112 for support of this research. 

\bibliographystyle{plain}
\bibliography{biblio}
\end{document}